\documentclass[aps, twocolumn]{revtex4-1}

\usepackage{times}
\usepackage{epsfig}
\usepackage{url}
\usepackage{soul}
\usepackage{graphicx}
\usepackage{graphics}
 \usepackage[usenames,dvipsnames]{xcolor}
\usepackage{amssymb,amsmath,amsthm,bbm}
\usepackage{latexsym}
\usepackage{subfigure}
\usepackage{wrapfig}
\usepackage[makeroom]{cancel}
\usepackage[justification=justified, small,bf]{caption}
\usepackage{framed}
\usepackage{float}
\usepackage[colorlinks={true},breaklinks]{hyperref}  
\hypersetup{
  colorlinks,
  citecolor=PineGreen,
  linkcolor=MidnightBlue,
  urlcolor=MidnightBlue}

\theoremstyle{definition}
\newtheorem{definition}{Definition}[section]

\newcommand{\tr}{{\rm tr\thinspace}}
\newcommand{\bra}[1]{\ensuremath{\left\langle{#1}\right\vert}}
\newcommand{\ket}[1]{\ensuremath{\left\vert{#1}\right\rangle}}

\newcommand{\expect}[1]{\ensuremath{\left\langle{#1}\right\rangle}}

\newcommand{\dg}{^{\dagger}}

\newcounter{aside}[section]

\renewcommand{\theaside}{\thesection.\arabic{aside}}

\usepackage[framemethod=TikZ]{mdframed}

\newcommand{\nn}{\nonumber}
\newcommand{\eg}{\emph{e.g.},~}
\newcommand{\ie}{\emph{i.e.},~}
\newcommand{\etal}{\emph{et al.}~}

\newcommand{\e}{\mathrm{e}}
\renewcommand{\d}{\mathrm{d}}
\renewcommand{\i}{\mathrm{i}}
\newcommand{\<}{\langle}
\renewcommand{\>}{\rangle}
\renewcommand{\vec}[1]{\boldsymbol{#1}}
\renewcommand{\a}{\hat{a}}
\newcommand{\adagger}{\hat{a}^{\dagger}}

\newcommand{\dis}{\displaystyle}

\newcommand{\UCB}{Department of Applied Mathematics, University of Colorado, Boulder, CO 80309-0526, USA}
\newcommand{\SNL}{Extreme-scale Data Science and Analytics,
Sandia National Laboratories, Livermore, CA 94550, USA}

\theoremstyle{definition} 
\theoremstyle{definition} \newtheorem*{Remark*}{Remark}

\usepackage{cleveref} 
\crefformat{equation}{Eq.~(#2#1#3)} 
\crefmultiformat{equation}{Eqs.~(#2#1#3)}{ and~(#2#1#3)}{, (#2#1#3)}{ and~(#2#1#3)}
\Crefformat{equation}{Equation~(#2#1#3)}
\crefformat{section}{Sec.~#2#1#3}
\Crefformat{section}{Section~#2#1#3}
\crefformat{figure}{Fig.~#2#1#3}
\Crefformat{figure}{Figure~#2#1#3}
\crefformat{aside}{Eg.~#2#1#3}
\Crefformat{aside}{Example~#2#1#3}
\crefmultiformat{aside}{Eg.~(#2#1#3)}{ and~(#2#1#3)}{, (#2#1#3)}{ and~(#2#1#3)}
\Crefmultiformat{aside}{Examples~(#2#1#3)}{ and~(#2#1#3)}{, (#2#1#3)}{ and~(#2#1#3)}
\crefformat{Remark}{Rem.~#2#1#3}
\Crefformat{Remark}{Remark~#2#1#3}

\setcitestyle{numbers,square}

\begin{document}

\title{Smoothing of Gaussian quantum dynamics for force detection}

\author{Zhishen Huang}
\email{zhishen.huang@colorado.edu}
\address{\UCB}

\author{Mohan Sarovar}
\email{mnsarov@sandia.gov}
\address{\SNL}

\date{\today}

\begin{abstract}
Building on recent work by Gammelmark \etal [Phys. Rev. Lett. 111, 160401 (2013)] we develop a formalism for prediction and retrodiction of Gaussian quantum systems undergoing continuous measurements. We apply the resulting formalism to study the advantage of incorporating a full measurement record and retrodiction for impulse-like force detection and accelerometry. 
\end{abstract}

\maketitle
 
\section{Introduction}\label{sec:intro}
Quantum sensing and metrology are rapidly maturing subfields of quantum information technology. Building on the historical precedent set by atomic clocks of using quantum systems to precisely measure quantities \cite{Goebel:2015wn}, there now exist a wide array of quantum sensors for tasks ranging from accelerometry to thermometry \cite{Degen:2017kw}. These applications motivate a closer examination of the techniques used to process the measurement records from such quantum sensors. In particular, a pertinent question is whether classical estimation algorithms are optimal for quantum sensors, since they may not take into account uniquely quantum phenomena such as measurement backaction \cite{Wiseman:2009vw}. This is especially of concern when measurements on the quantum system are continuous and weak, in which case the effects of backaction are non-negligible. In response to this question, a variety of techniques, that generally fall under the umbrella term of quantum filtering and estimation, have been developed over the past several decades.

The notion of \emph{smoothing} has recently been introduced into quantum estimation. Smoothing estimates some property of the quantum system at time $t^*$ using information in measurement record(s) up till that time, \emph{and} after this time \cite{Rhodes:1971iv}. This is in distinction to quantum filtering, which only uses the measurement record up till time $t^*$ to construct an estimate. Tsang first introduced smoothing in the quantum context for the purpose of estimating a classical signal based on measurements on a quantum system that is driven by the classical signal \cite{Tsang2009,Tsang:2009hi,Tsang:2010hf}. Tsang's smoothing approach has been demonstrated to be useful for several estimation problems where a classical signal of interest is tranduced by a quantum system, \eg \cite{Tsang:2011dq,Wheatley:2015jc}. 

Recently, Gammelmark \etal \cite{Gammelmark:2013co} developed a revised notion of a quantum state for a system undergoing continuous measurement that takes into account information in the measurement record at all times. As argued by Gammelmark \etal 
 this so-called \emph{past quantum state formalism} is a more direct generalization of smoothing in the classical context since it reconstructs a ``quantum state'' conditioned on past and future measurements, just like classical smoothing reconstructs a state. We note that Tsang had also suggested earlier that such a reconstruction of a past state is possible \cite[Sec. V]{Tsang:2009hi}, and that other interpretations of the notion of ``quantum state smoothing'' exist \cite{Guevara:2015di}. 
The predictions from the past quantum state formalism can be interpreted as quantum weak values since they provide estimates of observables based on post-selecting on a particular future measurement record \cite{Gammelmark:2013co, Wis-2002}. 

In this work we specialize the past quantum state formalism of Gammelmark \etal to continuously measured quantum systems that preserve Gaussian states. Such \emph{Gaussian dynamics} are relevant for harmonic systems (\eg coupled oscillators) where common state preparation protocols prepare Gaussian states and most dynamical processes and measurements are Gaussian. We investigate the extent to which smoothing using the past quantum state formalism can improve the performance a canonical estimation problem in the Gaussian harmonic oscillator context, namely, detection of external forces. 

We note that this specialization of the past quantum state formalism to Gaussian dynamics has also recently been independently developed by Zhang and M\o lmer \cite{Zhang:2017tw}\footnote{
Preliminary results presented in this paper appeared in ZH's internship conclusion report submitted to the Oak Ridge Institute for Science and Education (ORISE), on August 22, 2017.}. The dynamical equations we derive in section \ref{sec:gauss} are the same as the ones derived in Ref. \cite{Zhang:2017tw}, except for notational differences. However, there are two differences between the treatments that we wish to highlight: (i) in \cref{sec:app} we emphasize a different application from Zhang and M\o lmer, which requires inclusion of classical driving of Gaussian systems into the dynamical equations; and (ii) we derive equations of motion for the Gaussian \emph{information matrix} in \cref{sec:gauss}, which is critical for stable numerical simulation of the dynamics. 

The remainder of this article is organized as follows. In \cref{sec:smes} we review the formalism used to describe conditional states of continuously monitored quantum systems and the past quantum state formalism of Gammelmark \etal~ In \cref{sec:gauss} we present the description of linear dynamics of Gaussian states, and also derive the specialization of the past quantum state formalism to the Gaussian setting. Then in \cref{sec:app} we simulate the new dynamical equations and present an application of the formalism to impulse-like force detection. Finally, \cref{sec:disc} concludes with a brief discussion of future directions.

\section{Continuously measured quantum systems and the past quantum state formalism}
\label{sec:smes}
The state of quantum system undergoing Lindblad open system dynamics with one dissipation channel that is continuously monitored using a diffusive measurement (\eg homodyne monitoring) is described by the \emph{stochastic master equation} ($\hbar=1$) \cite{Jacobs_Review_Paper,wiseman2010quantum_Book}
\begin{align}
\label{forward_master_eqn}
\d\rho_t &= \mathcal{L}_0 \rho_t + \mathcal{L}_1 \rho_t, ~~~\textrm{with} \nn \\
\mathcal{L}_0 \rho_t &= -\i[\hat{H},\rho_t] \,\d t + \sum_{m=1}^M \mathcal{D}[\hat{L}_m]\rho_t \, \d t \nn \\
\mathcal{L}_1\rho_t &= \mathcal{D}[\hat{L}_0]\rho_t \d t + \sqrt{\eta}(\hat{L}_0\rho_t + \rho_t\hat{L}_0^\dagger) \, \d Y_t \nonumber
\end{align}
where $\d\rho_t = \rho_{t+\d t} - \rho_t$, and $\mathcal{D}[A]\rho \equiv A\rho A\dg - \frac{1}{2}\{A\dg A, \rho\}$. $\mathcal{L}_0$ represents the deterministic evolution of the system density matrix under the Hamiltonian $H$ and Lindblad operators $L_m$, $m\geq 1$, while $\mathcal{L}_1$ represents the evolution of the system due to the monitored channel ($m=0$). The stochastic quantity $\d Y_t$ is the increment in the observed measurement record, and is explicitly, 
\begin{align}
	\d Y_t = \sqrt{\eta}\expect{\hat{L}_0 + \hat{L}_0\dg}\d t + \d W_t,
\end{align}
where $\d W_t$ is a Wiener increment satisfying $E[\d W_s \d W_t] = \d t\cdot\delta(t-s)$, and $0\leq \eta \leq 1$ is the efficiency of the measurement. We note that this equation is linear in $\rho_t$ but does not preserve the trace of the density matrix. It is possible to write a nonlinear stochastic master equation that is explicitly trace-preserving \cite{Wis-1996}. 

The interpretation of the $\rho_t$ evolved by this equation is as the best estimate of the system conditioned on the information in the measurement record up till time $t$. In analogy to classical signal processing theory, this equation is sometimes called the \emph{quantum filtering equation}, which underscores the fact that it only takes into account information in the measurement record up till time $t$. In principle, it should be possible to also incorporate information from measurements after time $t$, if available, to refine the estimate of the state at time $t$. This is precisely what a \emph{smoothing} protocol does. 

Gammelmark \etal define a notion of smoothing in the quantum context by first defining backward-time evolution of a POVM \emph{effect} \cite{Kra-1983}, $E_t$, that is consistent with the forward-time evolution in \cref{forward_master_eqn}:
\begin{align}
\label{backward_master_eqn}
\d E_t = &\i[\hat{H},E_t]\,\d t + \sum_{m=0}^M \mathcal{D}\dg[\hat{L}_M]E_t\,\d t \nn \\
&+ (\hat{L}_0^\dagger E_t + E_t \hat{L}_0)\,\d Y_{t-\d t} 
\end{align}
where $\mathcal{D}\dg[A]\rho \equiv A\dg \rho A - \frac{1}{2}\{A\dg A, \rho\}$, $\d t$ is positive and $\d E_t = E_{t-\d t}-E_t$, propagates $E_t$ backward from some final time $t=T$ using the same measurement record $\d Y_t$ as in equation \eqref{forward_master_eqn}. The final condition for the effect is $E(T)=I$.
Gammelmark \etal prove that the forward-propagating density matrix and the backward propagating effect together define the best estimate of a measurement outcome at some intermediate time $t=t^*$ via the generalized Born rule
\begin{align}
p(m) = \frac{\mathrm{Tr}[\hat{\Omega}_m\rho(t^*)\hat{\Omega}^\dagger_m E(t^*)]}{\sum_k \mathrm{Tr}[\hat{\Omega}_k\rho(t^*)\hat{\Omega}^\dagger_k E(t^*)]},
\label{eq:gen_born_rule}
\end{align}
where the observable measured at time $t^*$ decomposes into POVM effects $\{\Omega_m\}$, with $\sum_m \Omega_m\dg \Omega_m = I$. For this reason, they define the tuple $\Xi_t = (\rho_t, E_t)$ as the \emph{past quantum state}, which provides better predictions (or more accurately, retrodictions) of properties conditioned on the information in the whole measurement record.

In the remainder of the paper we will assume that there are no dissipative channels in addition to the measurement channel. Therefore, $M=0$.

\section{Gaussian systems}
\label{sec:gauss}
Consider a system of $n$ harmonic oscillator modes with annihilation operators $\hat{a}_k$, $k=1,2,\cdots,n$, that satisfy the canonical bosonic commutation relations $[\hat{a}_k,\hat{a}_l^\dagger] = \delta_{kl}$. For each mode, we define the relationship between the annihilation/creation operators and the canonical quadrature operators via
\begin{align}
\begin{pmatrix}
\adagger_k\\
\a_k
\end{pmatrix} = \frac{1}{2}\begin{pmatrix}
1 & -\i\\
1 & \i
\end{pmatrix}\begin{pmatrix}
\hat{x_k}\\
\hat{p_k}
\end{pmatrix},\\
\begin{pmatrix}
\hat{x_k}\\
\hat{p_k}
\end{pmatrix} = \begin{pmatrix}
1 & 1\\
\i & -\i
\end{pmatrix}\begin{pmatrix}
\adagger_k\\
\a_k
\end{pmatrix},
\end{align}
and thus, $[\hat{x}_k,\hat{p}_l] = \i \delta_{kl}$.

\begin{definition}[Gaussian State]
A state, $\rho$, of $n$ harmonic modes is called a \textit{Gaussian state} if its Wigner function takes Gaussian form; \ie
\begin{equation}
\mathcal{W}[\rho](\vec{X}) = \frac{\e^{-\frac{1}{2}(\vec{X}-\expect{\vec{X}})^{\sf T} V^{-1}(\vec{X}-\expect{\vec{X}})}}{(2\pi)^n\sqrt{|V|}},
\end{equation}
where $\hat{\vec{X}} = \begin{pmatrix}
\hat{x}_1,
\hat{p}_1, 
\dots,
\hat{x}_n,
\hat{p}_n
\end{pmatrix}^{\mathsf T}$, $\expect{\vec{X}}_i = \expect{\hat{\vec{X}}_i}$, and $V$, the covariance matrix of the quadrature operators is defined as
\begin{align}
V_{nm} = \frac{\<\{\hat{\vec{X}}_n,\hat{\vec{X}}_m\}\>}{2} - \<\hat{\vec{X}}_n\>\<\hat{\vec{X}}_m\>
\end{align}
\end{definition}

\begin{figure*}
    \begin{subfigure}
        \centering
        \includegraphics[height=2.2in]{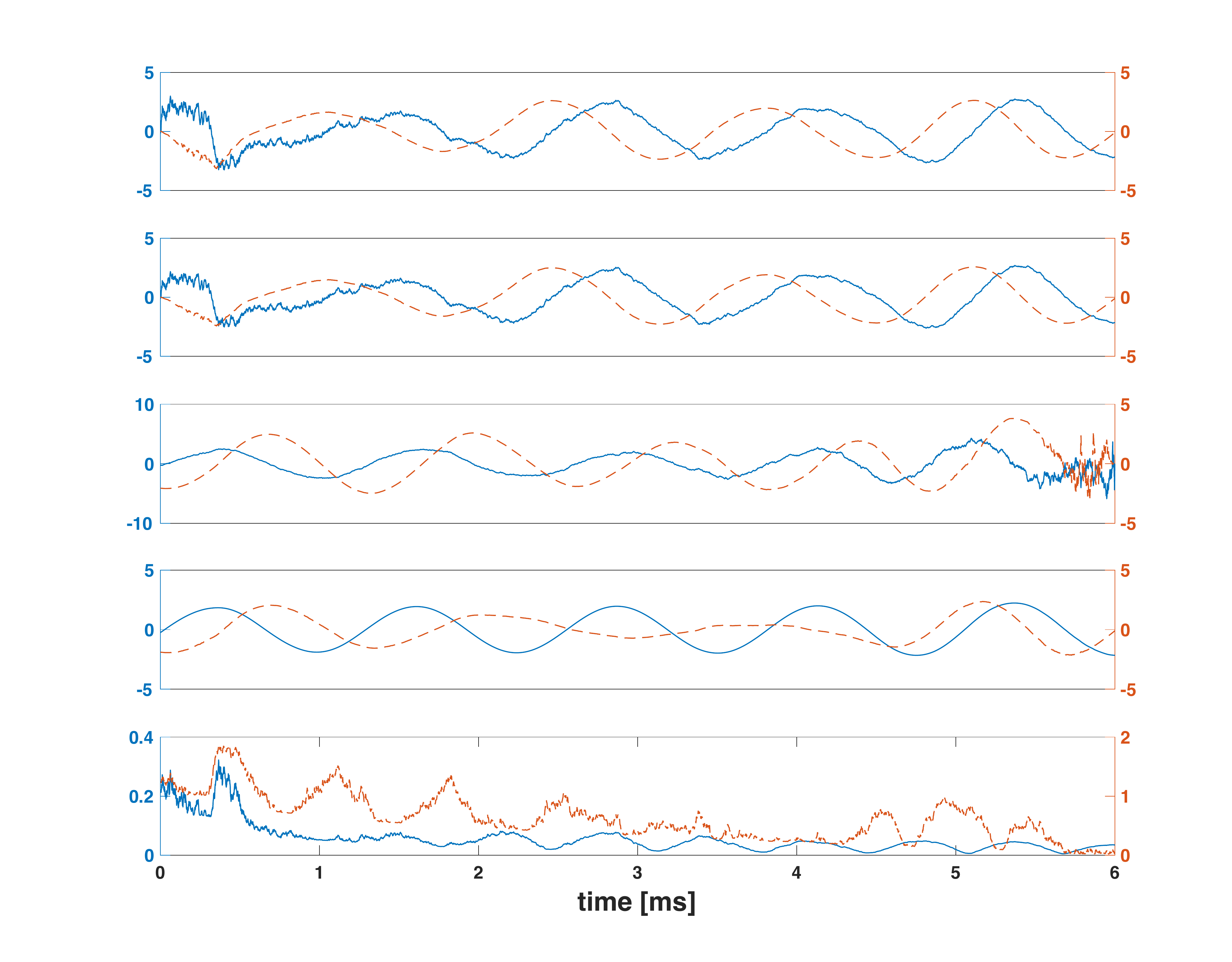}
    \end{subfigure}%
        \begin{subfigure}
        \centering
        \includegraphics[height=2.2in]{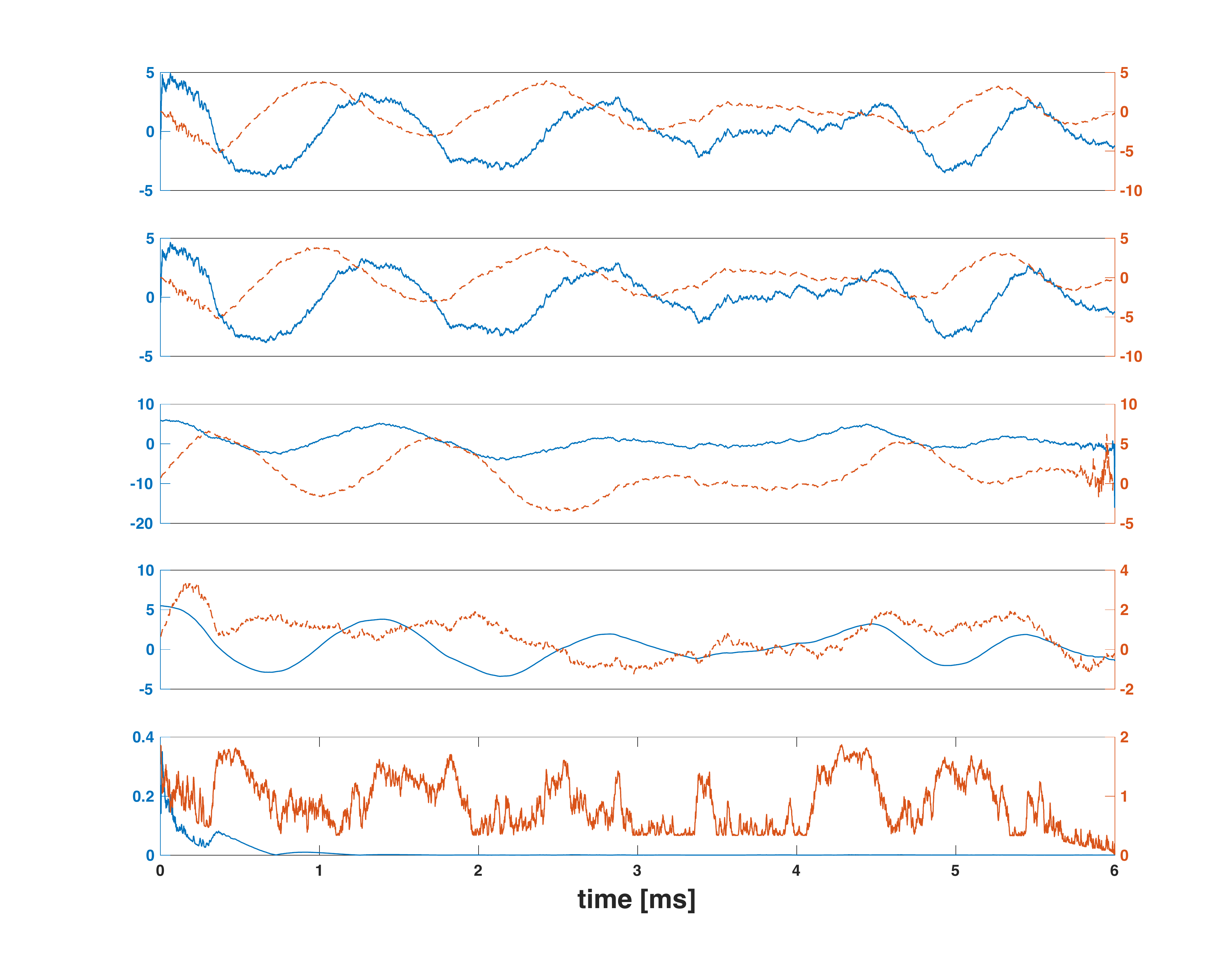}
    \end{subfigure}%
    \caption{Sample trajectories of the measured harmonic oscillator system with no external force ($u(t)=0$). The left (right) panel corresponds to $\kappa=0.1$kHz ($\kappa=2$kHz). In each panel, the axes, going from top to bottom, show $\expect{x}$ (blue, solid) and $\expect{p}$ (red, dashed) for the reference (R) system, forward evolving (F) system, backward evolving (B) system, and the smoothed (S) system. The final axis shows the difference at each time between the reference value of $\expect{x}$ and the value predicted by F (in blue) and the value predicted by S (red). The parameters used in the simulations are $n_R=5, n_F=3$.   \label{fig:trajs}}
\end{figure*}

The stochastic master equation in \cref{forward_master_eqn} preserves Gaussian states if (i) the Hamiltonian is quadratic in the canonical coordinates, and (ii) each $\hat{L}_m$ is linear in these coordinates \cite{Wis.Doh-2005, Wiseman:2009vw}. In this case we refer to the dynamics as linear, and the dynamical system as a linear system. More precisely, let the Hamiltonian of the $n$ modes take the following form:
\begin{align}
\hat{H} = \frac{1}{2}\vec{X}^{\sf T} G\vec{X}- \vec{X}^{\sf T} \vec{\Omega} B\vec{u}(t)
\end{align}
for some real, symmetric $2n\times 2n$ matrix $G$, and some time-dependent classical driving on the system, $\vec{u}(t)$. The matrix $B$ is real and has dimensions $2n \times m$, where $m$ is the number of modes that are subject to driving/forcing. $\vec{\Omega}$ is the $(2n)\times(2n)$ symplectic form $\dis \bigoplus_{k=1}^n\begin{pmatrix}
0 & 1\\
-1 & 0
\end{pmatrix}$.  Furthermore, since the measurement operator is linear in the canonical coordinates, we can write it as, $\hat{L}_0 = \tilde{C}\vec{X}$ for some $1\times 2n$ matrix $\tilde{C}$.

Using $\d\expect{\hat{z}}_t = \tr(\hat{z}\d\rho_t)$ for any operator $\hat{z}$, we can equivalently express \cref{forward_master_eqn} in terms of dynamical equations for the means and covariance matrix of the time-dependent Gaussian state \cite{Wis.Doh-2005}:
\begin{flalign}
\label{dynam_forw}
&\d \<\vec{X}\>_t = [A\<\vec{X}\>_t+B\vec{u}_t]\,\d t + \sqrt{\eta}(V_tC^{\sf T}+\Gamma^{\sf T})\,\d W_t\nn \\
&\frac{\d V_t}{\d t} = AV_t + V_tA^{\sf T} + D - \eta (V_tC^{\sf T}+\Gamma^{\sf T})(CV_t+\Gamma)
\end{flalign}
where $A = \vec{\Omega}(G+\mathfrak{Im}[\tilde{C}^\dagger\tilde{C}])$, $C = 2\mathfrak{Re}[\tilde{C}]
$, $D = \vec{\Omega}\mathfrak{Re}[\tilde{C}^\dagger\tilde{C}]\vec{\Omega}^{\mathsf T}$, $\Gamma = -\mathfrak{Im}[\tilde{C}] \vec{\Omega}^{\mathsf T}$.
Here $^{\mathsf T}$ denotes matrix transpose, $^\dagger$ denotes Hermitian conjugate, $^*$  denotes complex conjugate, and $\mathfrak{Re/Im}$ denotes taking element-wise real/imaginary parts of a matrix. 
The initial conditions for these equations are the mean and covariance matrix for the initial Gaussian state.

Of course, the above formulation can be generalized to multiple measurement channels \cite{Wis.Doh-2005}, however we will not need this generalization in the following and hence we restrict ourselves to this simpler case.

\subsection{Gaussian formulation of backward evolution}
Just as with states, one can also define Gaussian measurements. Common examples of Gaussian measurements are homodyne and heterodyne measurements in optics. Gaussian measurements have POVM effects that can be represented by Wigner functions in Gaussian form \cite{Fiurasek:2007do}, \ie
\begin{equation}
\mathcal{W}[E_{\vec{Y}}](\vec{X}) = \frac{\e^{-\frac{1}{2}(\vec{X}-\vec{Y})^{\sf T} U^{-1}(\vec{X}-\vec{Y})}}{(2\pi)^n\sqrt{|U|}},
\end{equation}
where $\vec{Y}$ is a $2n\times 1$ vector of scalars that parameterize the POVM effect (\ie the outcomes corresponding to the measurement outcome represented by the effect). The covariance matrix of the effect is denoted $U$ to clearly distinguish it from the covariance matrix of a state (which we will always denote $V$).

As we did for state dynamics, using $\d\expect{\hat{z}}_t = \tr(\hat{z} \d E_t)$, we can translate the backward evolution of the POVM effect prescribed by \cref{backward_master_eqn} into dynamical equations for the ``means'' and covariance matrix describing the POVM effect:
\begin{flalign}
\label{dynam_back}
\d\vec{Y}_t = 
&-[A\vec{Y}_t + B\vec{u}_t]\,\d t+ \sqrt{\eta}(U_tC^{\sf T} - \Gamma^{\sf T})\,\d W_{t-dt}  \nn \\
\frac{\d U_t}{\d t} = &-AU_t - U_tA^{\sf T}  + D - \eta(U_tC^{\sf T}-\Gamma^{\sf T})(CU_t-\Gamma) 
\end{flalign}
These equations describe how to back-propagate these quantities from the final time $T$ to any intermediate time; \ie $\d \vec{Y}_t = \vec{Y}_{t-\d t}-\vec{Y}_t$. 

The initial (actually, final) conditions for these equations must correspond to the choice $E(T)=\hat{I}$, which raises an issue. The identity operator can only be approximated by a Gaussian state, since it corresponds to $\vec{Y}_T=0$ and $U_T=\textrm{diag}(\infty)$. We have found empirically that choosing $U_T=\textrm{diag}(\nu)$ for a large $\nu$ is often a suitable approximation that skirts this issue, but one can also obtain a more elegant solution by propagating the inverse of $U_t$ instead of $U_t$ itself. This is common practice in the literature on Kalman filtering, where $P_t \equiv U_t^{-1}$ is called the \emph{information matrix}. Using $\partial_t(U_t^{-1} U_t)=0$ and the product rule, one can derive from \cref{dynam_back},
\begin{align}
	\frac{\d P_t}{\d t} =& P_t A + A^{\sf T} P_t - P_t D P_t + \eta (C^{\sf T} - P_t\Gamma^{\sf T})(C - \Gamma P_t)
\end{align}
with initial condition $P_T=0$. 
We find that propagating this equation poses no numerical instability issues.

Now that we have forward and backward evolution equations for Gaussian parameterizations of the density matrix and POVM effect, the final ingredient necessary for a Gaussian formulation of the past quantum state formalism of Gammelmark \etal is the Gaussian equivalent of the generalized Born rule in \cref{eq:gen_born_rule}. Of course, one could simply construct the density matrix and effect at time $t^*$ from their Gaussian parameterizations and apply \cref{eq:gen_born_rule}, however it is more efficient to avoid explicit reconstruction of these operators. In order to do this, we will restrict ourselves to predicting probabilities of projecting onto multimode coherent states at the intermediate time $t^*$. That is, we will assume that the POVM elements $\Omega_m$ in \cref{eq:gen_born_rule} are $\hat{\Omega}_m \rightarrow \hat{\Omega}_\gamma = \ket{\gamma}\bra{\gamma}$ for some $n$-mode coherent state $\gamma$. Hence, \cref{eq:gen_born_rule} describes a \emph{probability density}
\begin{align}
f_{\rm smoothed}(\gamma, t^*) &= \frac{\mathrm{Tr}[\hat{\Omega}_\gamma\rho(t^*)\hat{\Omega}^\dagger_\gamma E(t^*)]}{\int \mathrm{Tr}[\hat{\Omega}_\alpha\rho(t^*)\hat{\Omega}^\dagger_\alpha E(t^*)]	\,\d^{2n}\alpha } \nn \\
&= \frac{\mathcal{Q}_{\rho,t^*}(\gamma)\mathcal{Q}_{E,t^*}(\gamma)}{\int \mathcal{Q}_{\rho,t^*}(\alpha)\mathcal{Q}_{E,t^*}(\alpha) \, \d^{2n}\alpha},
\label{eq:gen_bor_rule_q}
\end{align}
where $\mathcal{Q}_A(\alpha) = \frac{1}{\pi^n}\bra{\alpha}A\ket{\alpha}$ is the $n$-mode Husimi Q-function of operator $A$ (again, $\alpha$ is an $n$-mode coherent state). For a Gaussian operator, the Q-function takes Gaussian form \cite{Gar.Zol-2004} and is related to the Wigner function via the integral transform: 
$\mathcal{Q}_A(\alpha) = \frac{2}{\pi}\int \mathcal{W}(\beta)\e^{-2|\alpha - \beta|^2}\,\d^{2n}\beta$. 

\cref{eq:gen_bor_rule_q} allows efficient calculation of a smoothed probability density in terms of the Gaussian parameters that are propagated by \cref{dynam_back,dynam_forw}. One can contrast $f_{\rm smoothed}(\gamma, t^*)$ against what this probability density would be if one only relies on measurements prior to $t^*$, \ie the output of a Gaussian filter. This is $f_{\rm filtered}(\gamma, t^*) = \mathcal{Q}_{\rho, t^*}(\gamma)$. Thus, the smoothing applies a Gaussian smoothing kernel 
formed from information in future measurements.
Hence in the Gaussian context, the smoothed probability density estimate is a Gaussian blur of the filtered estimate. To see the effect of this, consider the smoothed estimate of the probability distribution obtained by a homodyne measurement of one of the quadratures of a single mode, \ie $\hat{\Omega}_\gamma = \ket{x}\bra{x}, x\in \mathbb{R}$. By noting that \cref{eq:gen_bor_rule_q} prescribes a multiplication of two Gaussian functions, we can obtain explicit forms for the mean at time $t^*$, and the variance of this estimate:
\begin{align}
\<x\>_S(t^*) &= \frac{U_{11}}{V_{11}+U_{11}}\expect{x}_F(t^*) + \frac{V_{11}}{V_{11}+U_{11}}\expect{x}_E(t^*) \nn \\
\sigma^2 \big(\<x\>_S(t^*) \big) &= \frac{1}{\frac{1}{V_{11}}+\frac{1}{U_{11}}}, \nn
\end{align}
where all the covariance matrix elements are also evaluated at time $t^*$ but we omit this index for notational simplicity.

\section{Application: force detection}
\label{sec:app}
In this section we apply the above formalism to the canonical problem of force detection using a harmonic system. In particular, we develop a practical protocol for detection of impulsive forces. We note that Tsang's smoothing formalism has been applied to similar physical context, but typically to estimate spectra of continuous driving signals, \eg \cite{Tsang:2011dq,Wheatley:2015jc}. In the following, we place an emphasis on detecting the presence and arrival-time of impulse-like forces.

Consider a single harmonic mode undergoing free evolution, driving by some unknown time-dependent force, $u(t)$, and weak, continuous measurement of its position. This system is described by the Hamiltonian and measurement operator:
\begin{align}
\hat{H} &= \frac{\omega_a}{4}(\hat{x}^2+\hat{p}^2) + u(t)\hat{x} \nn \\
&= \vec{X}^{\sf T} 
\begin{pmatrix}
\omega_a/4 & 0 \\
 0 & \omega_a/4 
\end{pmatrix} 
\vec{X} + u(t) 
\begin{pmatrix}
1 & 0 
\end{pmatrix}
\vec{X} \nn \\
\hat{L}_0 &= \sqrt{2\kappa}\hat{x} \nn \\
&= \begin{pmatrix}
\sqrt{2\kappa} & 0 
\end{pmatrix}
\vec{X},
\end{align}
where in the second line of each term we have written the Hamiltonian and measurement operator in terms of the matrices in the linear systems theory, and $\vec{X} = (\hat{x}, \hat{p})^{\sf T}$. We assume the measurement is efficient, and therefore set $\eta=1$. We use natural units to measure length, \ie in units of $\sqrt{1/m\omega_a}$ where $m$ is the mass of the oscillator, and hence $u(t)$ has units of $1/s$. In addition, we set $\omega_a= 10$kHz for concreteness. 

\begin{figure}
        \includegraphics[height=2.5in]{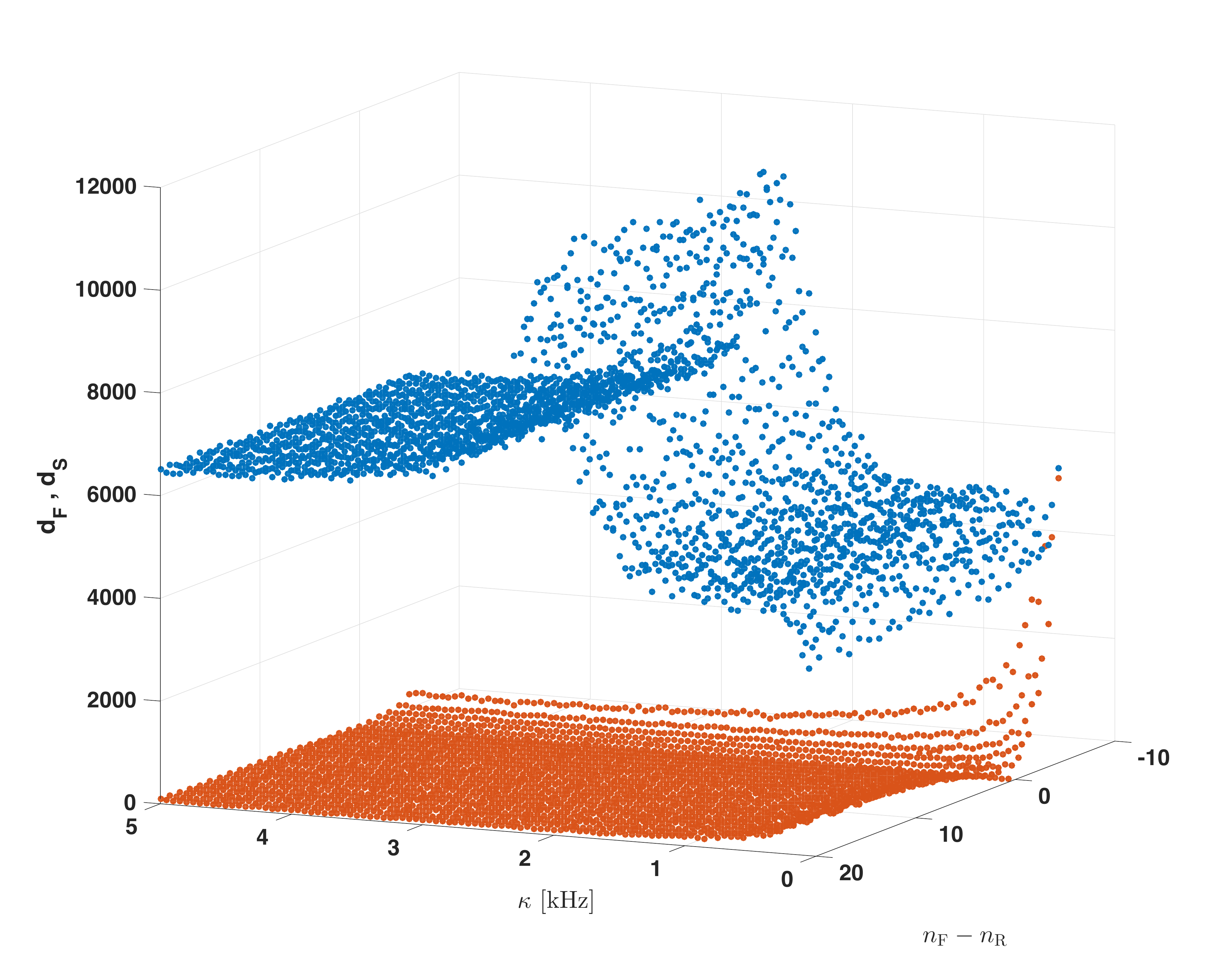}
    \caption{Accuracy of reconstruction of probability distribution for oscillator position as a function of the measurement strength ($\kappa$) and error in the initial state of the filter system (F). The other parameters used are $n_R=5$. The red (blue) surface is $d_F$ ($d_S$) \label{fig:state_recon_sweep}}
\end{figure}

We simulate these dynamics for a reference system (R), which produces a system evolution trajectory and the measurement current
\begin{align}
\d I(t) = \expect{\hat{x}}_R(t)\d t + \frac{\d W(t)}{\sqrt{8\kappa}},
\end{align}
where the expectation value is under the state of the reference system, $\expect{\hat{x}}_R(t) \equiv (\vec{X}^R_t)_1$, and $\d W(t)$ are independent Wiener increments. The initial state of the reference system is assumed to be a thermal state, \ie
\begin{align}
\vec{X}^R(0) = (0, 0)^{\sf T}, ~~~~ V^R(0) = (2\bar{n}_R+1)I_2	
\end{align}

\begin{figure}
    \begin{subfigure}
        \centering
        \includegraphics[height=2.2in]{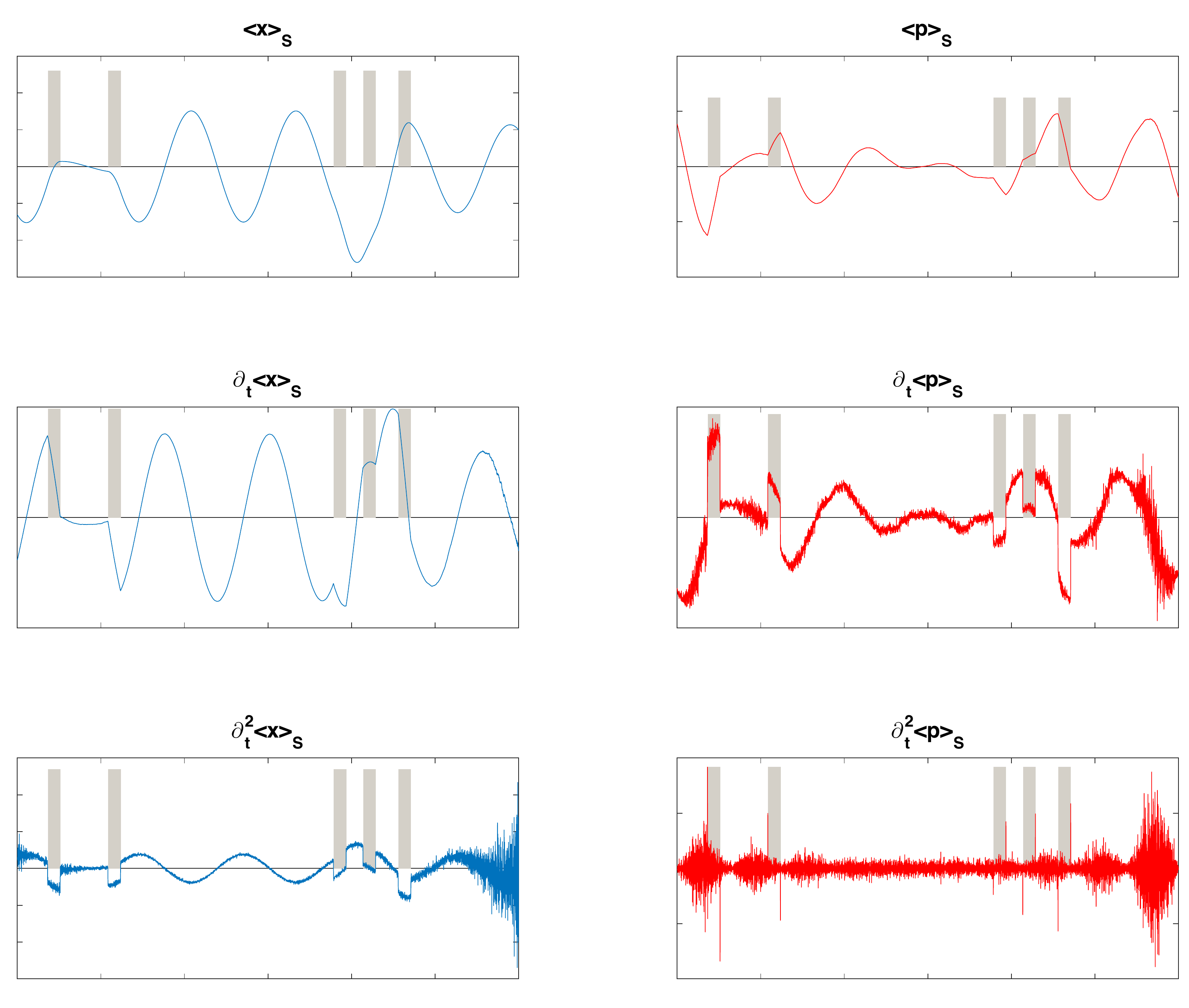}
    \end{subfigure}%
    \quad\quad\quad
    \begin{subfigure}
        \centering
        \includegraphics[height=2.2in]{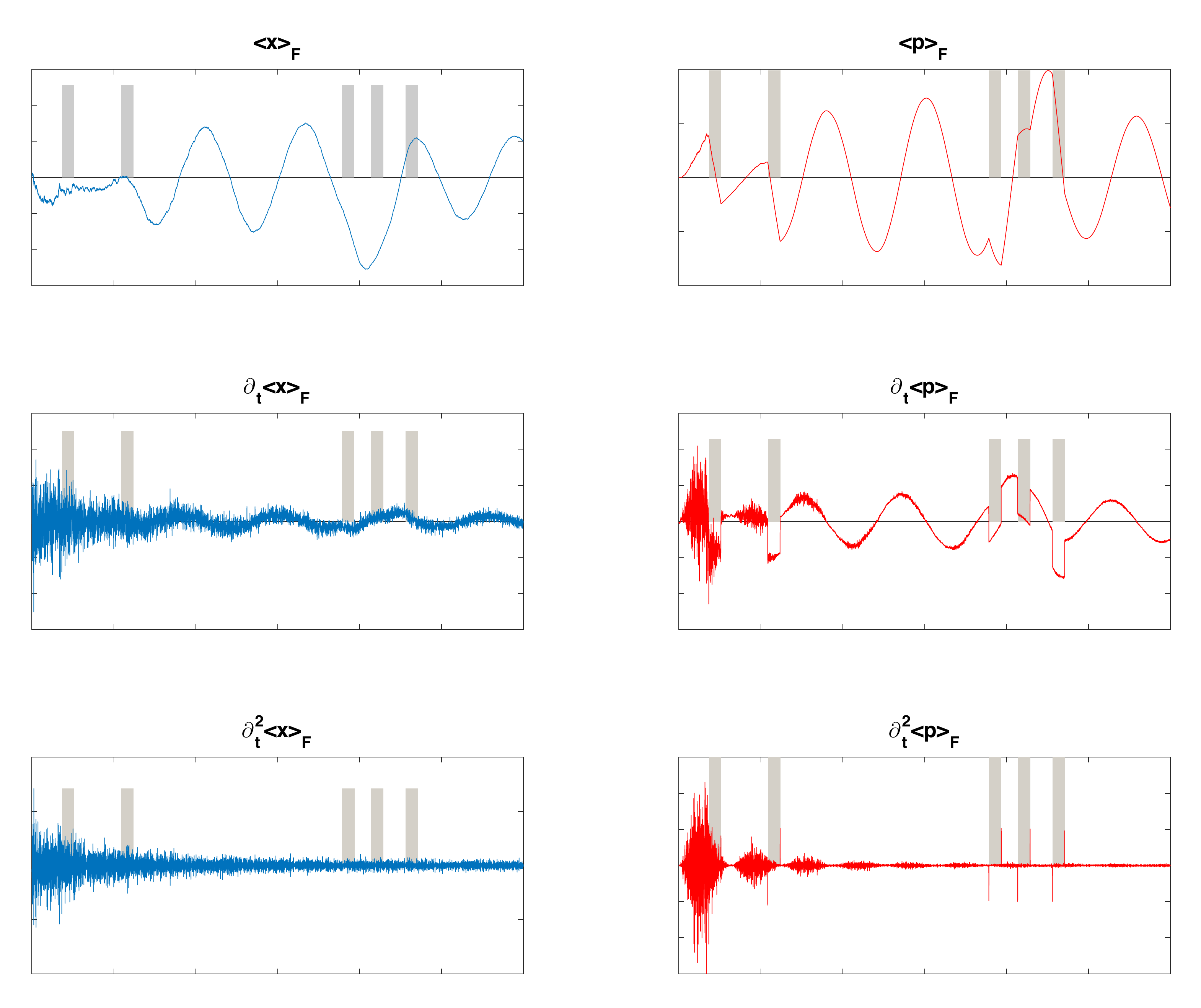}
    \end{subfigure}%
    \caption{Sample trajectories of a measured harmonic oscillator system subject to a series of impulse-like external forces (indicated by gray rectangles in all figures). The top three (bottom three) rows show coordinate predictions from the smoothed (forward) system. The left column (with blue lines) show $\expect{x}$ and its first and second derivatives for each system, while the right column (with red lines) show $\expect{p}$ and its first and second derivatives for each system. The parameters used in the simulations are $\kappa=0.1$kHz, $n_R=5, n_F=3, s=50\textrm{kHz}, w=0.15\textrm{ms}$.  
    \label{fig:impulse_trajs}}
\end{figure}

Then we use this ``experimental'' measurement current to drive the evolution of a Gaussian state evolving forward (F) in time according to \cref{dynam_forw}, and an effect matrix for a Gaussian measurement evolving backward (B) in time according to \cref{dynam_back}. The only quantity observed from the reference system is the measurement current, and therefore the innovations that drive systems F and B are formed as:
\begin{align}
	\d W^{\textrm{F}}(t) &= (\d I(t) - \expect{\hat{x}}_F(t) \,\d t)\sqrt{8\kappa} \nn \\
	\d W^{\textrm{B}}(t) &= (\d I(t) - \expect{\hat{x}}_B(t) \, \d t)\sqrt{8\kappa},
\end{align}
where $\expect{\hat{x}}_F(t) \equiv (\vec{X}^F_t)_1$ is the first component of the mean vector of the forward evolved system, and $\expect{\hat{x}}_B(t) \equiv (\vec{Y}^B_t)_1$ is the first component of the mean vector of the backward evolved system.

The initial state of the forward evolved system is also a thermal state, but we allow it to differ from the reference system initial state, \ie $\bar{n}_F \neq \bar{n}_R$. This accounts for any possible error in estimating the initial state of the system from which the measurement current is recorded. 

Finally, we form a smoothed system (S) by combining the predictions from F and B at each time $t$, according to the smoothed probability density \cref{eq:gen_bor_rule_q}.

\begin{figure*}
	\begin{subfigure}
	(a)
		\includegraphics[height=1.5in]{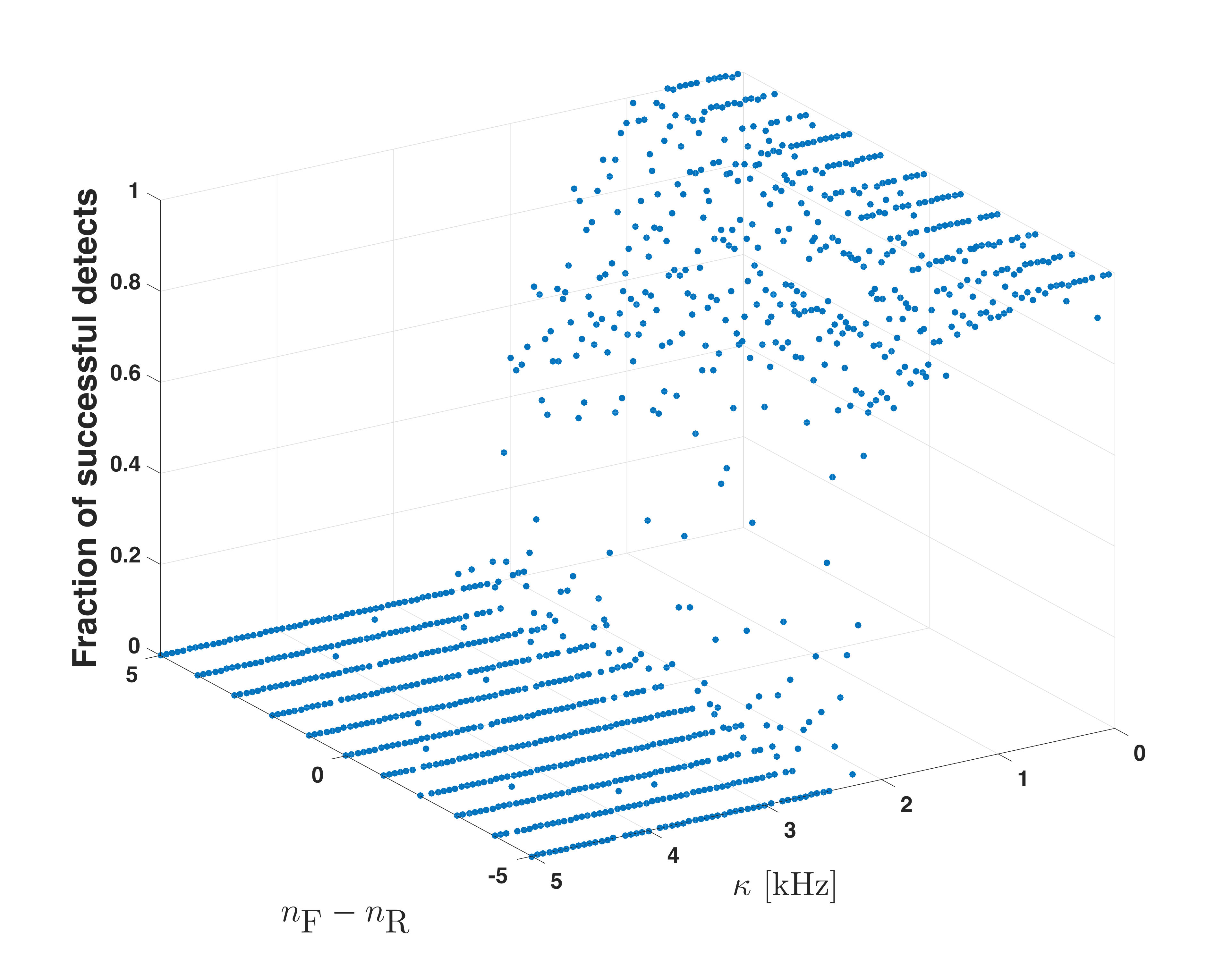}
	\end{subfigure}
	\begin{subfigure}
	(b)
		\includegraphics[height=1.5in]{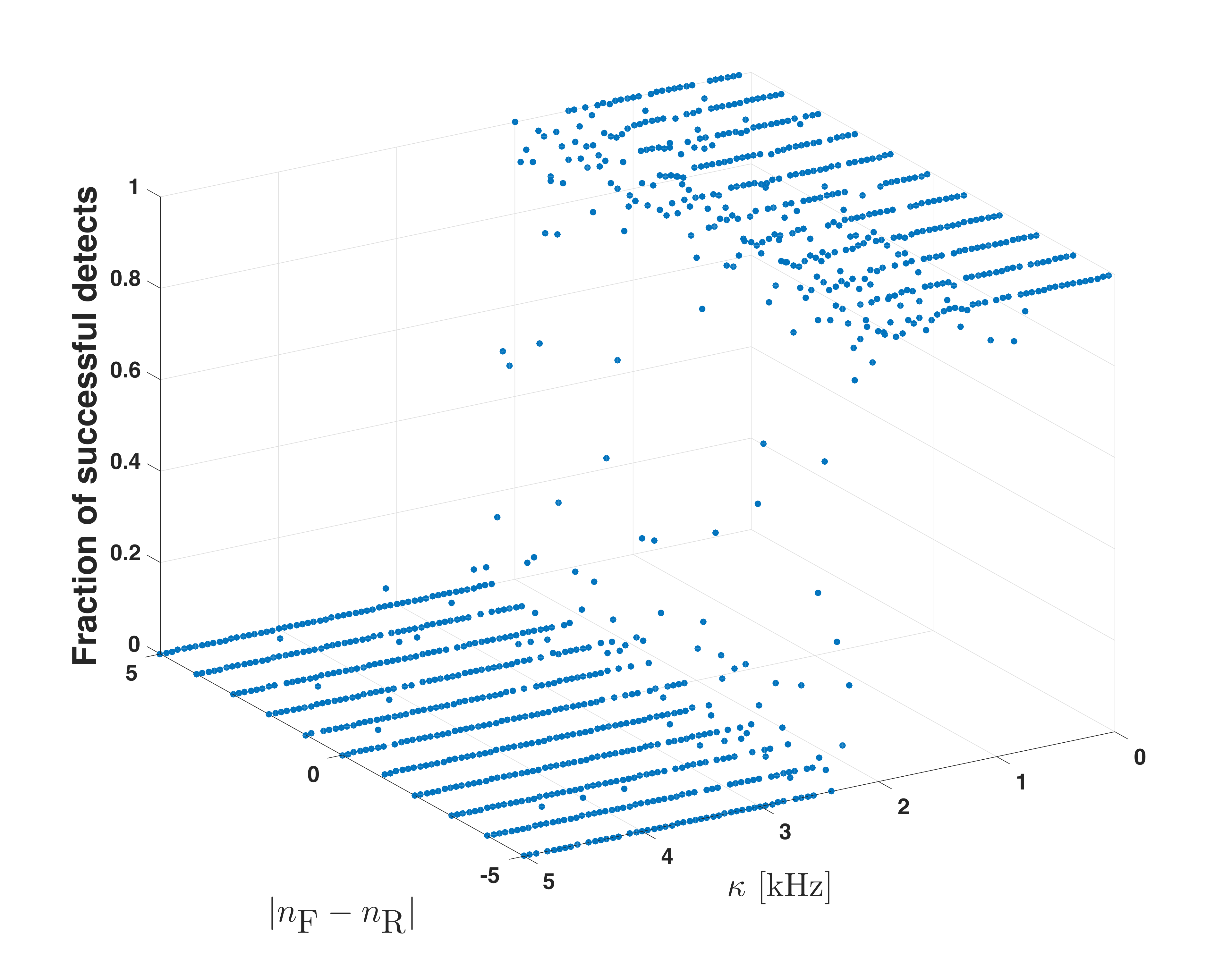}
	\end{subfigure}
	\begin{subfigure}
	(c)
		\includegraphics[height=1.5in]{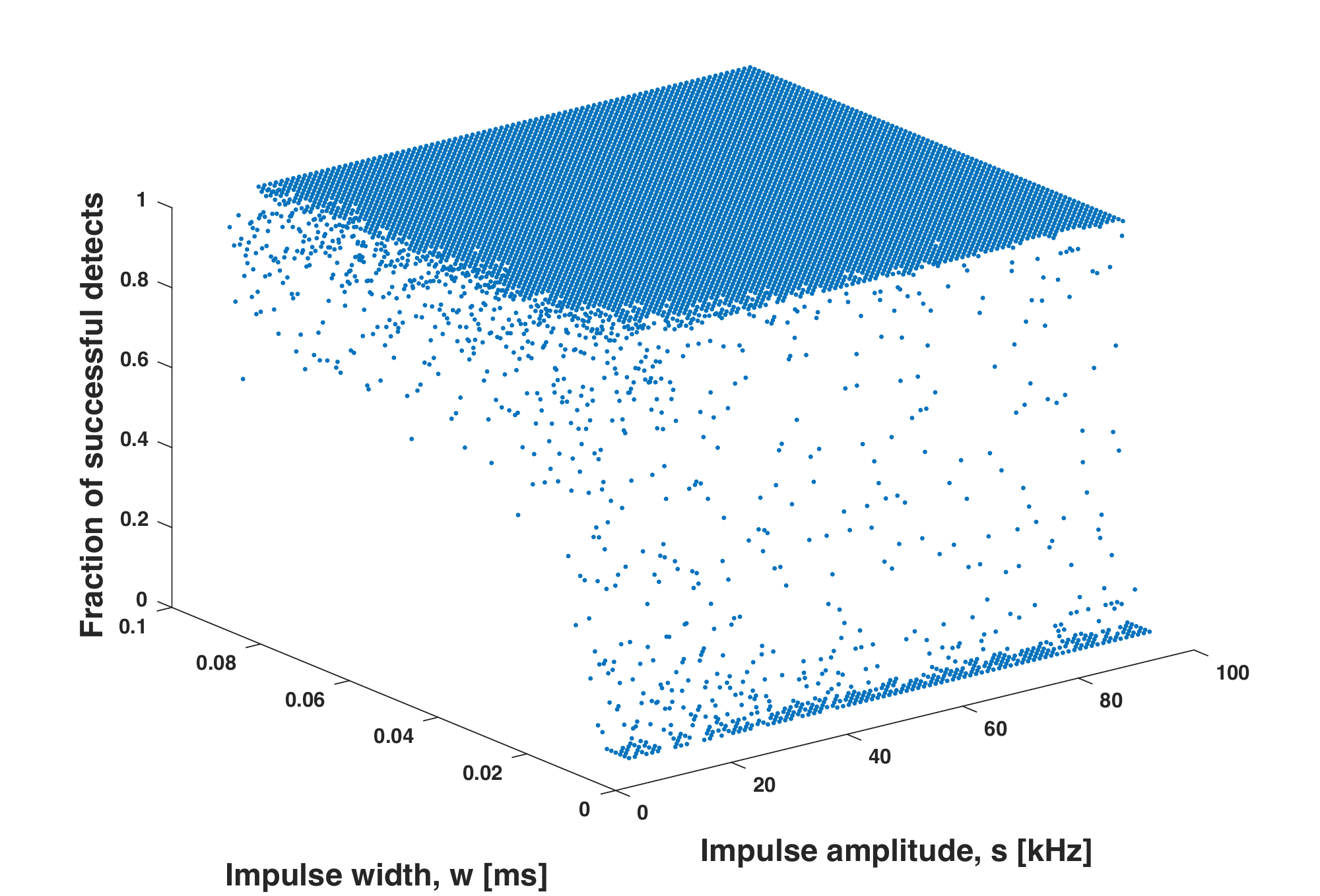}
	\end{subfigure}
	\caption{Accuracy of force-detection protocol as a function of system parameters. All plots show the fraction of successful detections of impulse-like forces. The first two plots sweep over the measurement strength ($\kappa$) and error in initial state estimate (($n_R$ is kept at 5, while $n_F$ is varied), while the last plot sweeps over impulse amplitude and width. At each parameter combination we ran $N=20$ simulations, each with $n_i=5$ randomly placed impulses within the time interval $[0,6]$ ms for each run, and counted the number of times the protocol successfully identified an impulse. The remaining parameters in each plot are: (a) $s=50$kHz and $w=0.15$ms, (b) $s=100$kHz and $w=0.15$ms, (c) $n_{\rm R}=n_{\rm F}=5, \kappa=0.1$kHz.\label{fig:impulse_sweep}}
\end{figure*}

In \cref{fig:trajs} we show some example trajectories for all four systems (R, F, B, S), for two values of measurement strength, $\kappa$, when the force is absent ($u(t)=0, \forall t$). We see from this figure that while the predictions from the system S are much smoother than the predictions from F, the smoothed estimate of the mean values of the oscillator are not as accurate as the filtered values produced by the F system, especially for larger $\kappa$. To explore the impact of the system parameters more systematically, we sweep over the measurement strength and the initial state mismatch in \cref{fig:state_recon_sweep} and evaluate the accuracy of the reconstructions produced by the systems F and S at each parameter combination. The accuracy is determined by calculating an approximation of the total variation distance of the probabilities for the position coordinate produced by each system over all times, \ie
\begin{align}
	d_F &= \sum_{t} \int_{\mathbb{R}} |\mathbb{P}^F_t(x) - \mathbb{P}^R_t(x)| \, \d x \nn \\
	d_S &= \sum_{t} \int_{\mathbb{R}} |\mathbb{P}^S_t(x) - \mathbb{P}^R_t(x)| \, \d x, \nn\end{align}
	where $\mathbb{P}^{R/F/S}_t(x)$ is the probability distribution for the $x$ coordinate at time $t$ predicted by the R/F/S system. This figure shows that the accuracy of the smoothed estimate is nearly always inferior to the filtered estimate, unless (i) the error in the initial state of the F system is large, and (ii) the measurement strength is very weak. In effect, due to the weakness of the measurement in this regime, the filter cannot recover enough information from the measurement record over the simulated time period to compensate for the error in initial state estimate. However, incorporating the information from the backward evolution increases the accuracy at later portions of the time window.
We see similar behavior if we examine the accuracy of predictions about the momentum coordinates as well (not presented here).

Despite smoothing showing no clear advantage over filtering for the task of recovering the state of the system (except in a small parameter regime), we can try to take advantage of the fact that the predictions of system S are much smoother than the corresponding predictions from the filter. Smooth trajectories enable one to define derivatives that are better behaved, and this fact motivates an impulsive force detection protocol; impulsive forces lead to sudden changes in the coordinates of the oscillator, and perhaps a derivative based algorithm could identify such events. \cref{fig:impulse_trajs} shows sample trajectories of coordinate predictions and their derivatives when the system evolves under impulse-like forces, \ie $u(t) = \sum_k \sqcap^s_w(t_k)$, where $\sqcap^s_w(t_k)$ is a square pulse of width $w$ and height $s$ centered at $t_k$. As expected, derivatives of the smoothed predictions (S) reliably indicate the location of the impulse-like force, while the derivatives of the filtered predictions (F) are often too noisy.

Informed by these observations we define a force-detection protocol that identifies impulse-like forces by looking for discontinuities in $\dis \frac{\d ^2\<x\>_S}{\d t^2}$. We declare the presence of a force if the discontinuity is larger than a given value (threshold detection, see Appendix for details). In \cref{fig:impulse_sweep} we examine the effectiveness of this protocol as a function of the system parameters by sweeping over values of measurement strength, error in estimate of initial state for the F system, and the parameters of the impulse-like force, ($s,w$). 
Figs. \ref{fig:impulse_sweep}(a) and \ref{fig:impulse_sweep}(b) show that the impulse-force detection using the smoothed predictions is more effective for weak measurements, and in fact it can achieve $100\%$ accuracy for very small $\kappa$. For large values of $\kappa$ the protocol becomes less effective because the measurement induced decoherence and projection dynamics dominate the effect of the external force and hence the signature of the impulse-like force is weakly imprinted in the coordinate trajectories. 
In addition, these figures show that the protocol has a weak dependence on the error in the initial state estimate -- it remains robust despite large errors in the estimate of the initial state. Finally, \cref{fig:impulse_sweep}(c) shows that for small values of $\kappa$, the detection protocol remains accurate for a wide range of impulses properties (width and amplitude). Only when the impulse becomes very weak $s<15\rm{kHz}$ or very short $w<0.02$ms, does the success rate diminish. 

\section{Discussion}
\label{sec:disc}
We have adapted the past quantum state formalism of Gammelmark \etal \cite{Gammelmark:2013co} to the setting of Gaussian quantum states preserved by linear dynamics. This description is especially relevant for experimental platforms such as nanomechanical resonators \cite{Reg.Teu.etal-2008a} and trapped ultracold atoms \cite{Spethmann:2015dl}, whose motional modes are often well-approximated by Gaussian states. These platforms have been proposed as good candidates for engineering force-detectors and accelerometers operating at the quantum limit. Hence, we have studied the benefits of smoothing Gaussian dynamics via the past quantum state formalism in order to detect impulse-like forces. 

All the simulations presented above consider ideal dynamics. A direction for future work is to understand the performance of the past quantum state formalism for smoothing Gaussian dynamics in the presence of (i) measurement inefficiencies, and (ii) additional environmental decoherence channels.

\begin{acknowledgements}
	We wish to thank Dan Stamper-Kurn, Jonathan Kohler, Justin Gerber and Emma Dowd for several useful discussions on the topics of filtering and smoothing in cold-atom experiments, and with Bengt Fornberg on signal processing.
	ZH was supported by NSF Mathematical Sciences Graduate Internship during part of this work. 
	Sandia National Laboratories is a multimission laboratory managed and operated by National Technology and Engineering Solutions of Sandia, LLC., a wholly owned subsidiary of Honeywell International, Inc., for the U.S. Department of Energy's National Nuclear Security Administration under contract DE-NA-0003525.
\end{acknowledgements}

\bibliography{gaussian_retrodiction.bib}

\appendix
\section{Threshold detection protocol}

From figure \ref{fig:impulse_trajs}, we can see that the influence of the noise is significant near the end of the time range, even for the smoothed trajectories. As a result, detecting a discontinuity in the second order derivative of $\<x\>$, by calculating the third order derivative is not generally feasible. \\

\begin{figure}
        \centering
        \includegraphics[height=1.8in]{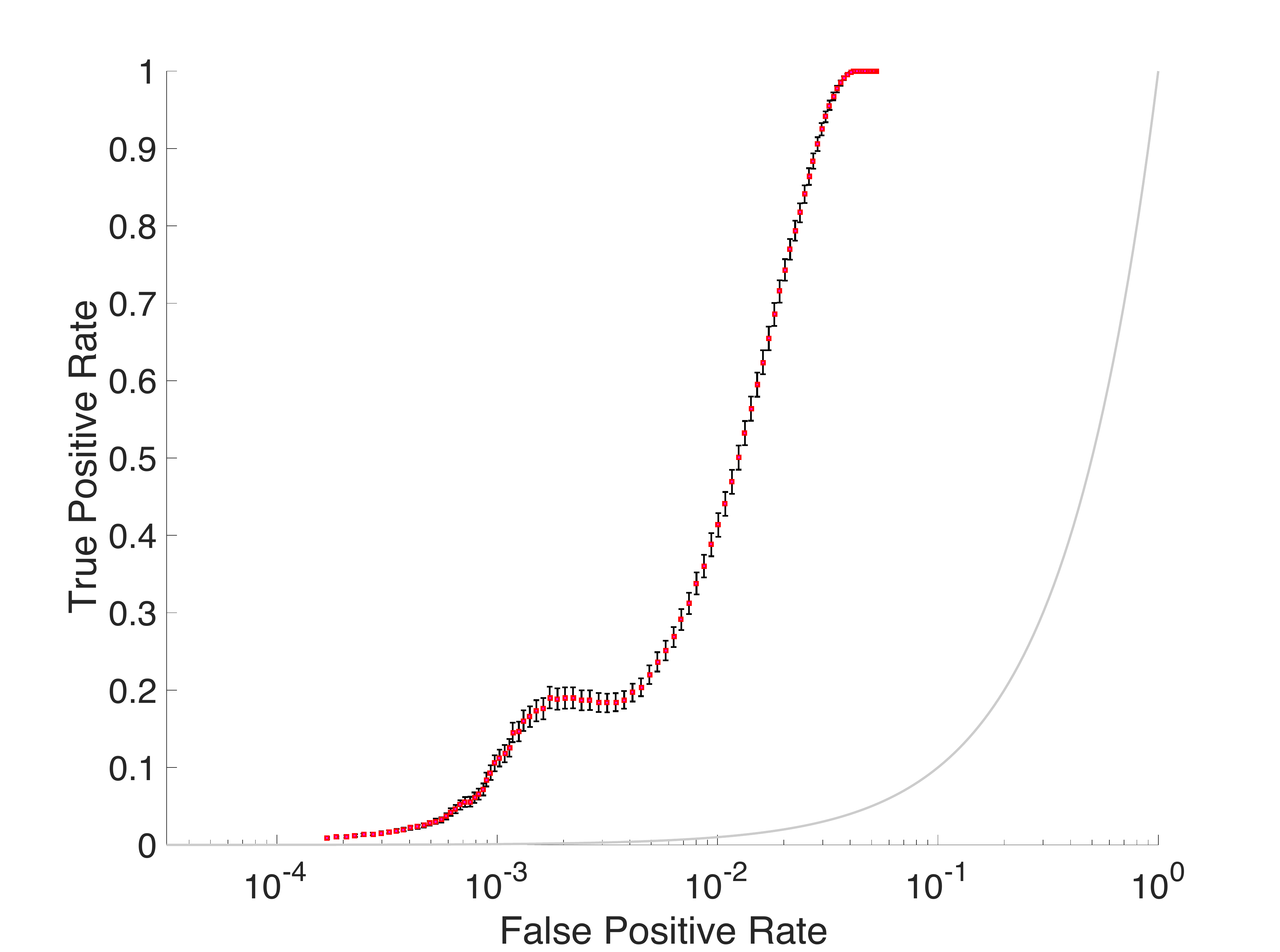}\\
        (a)\\
        \includegraphics[height=1.8in]{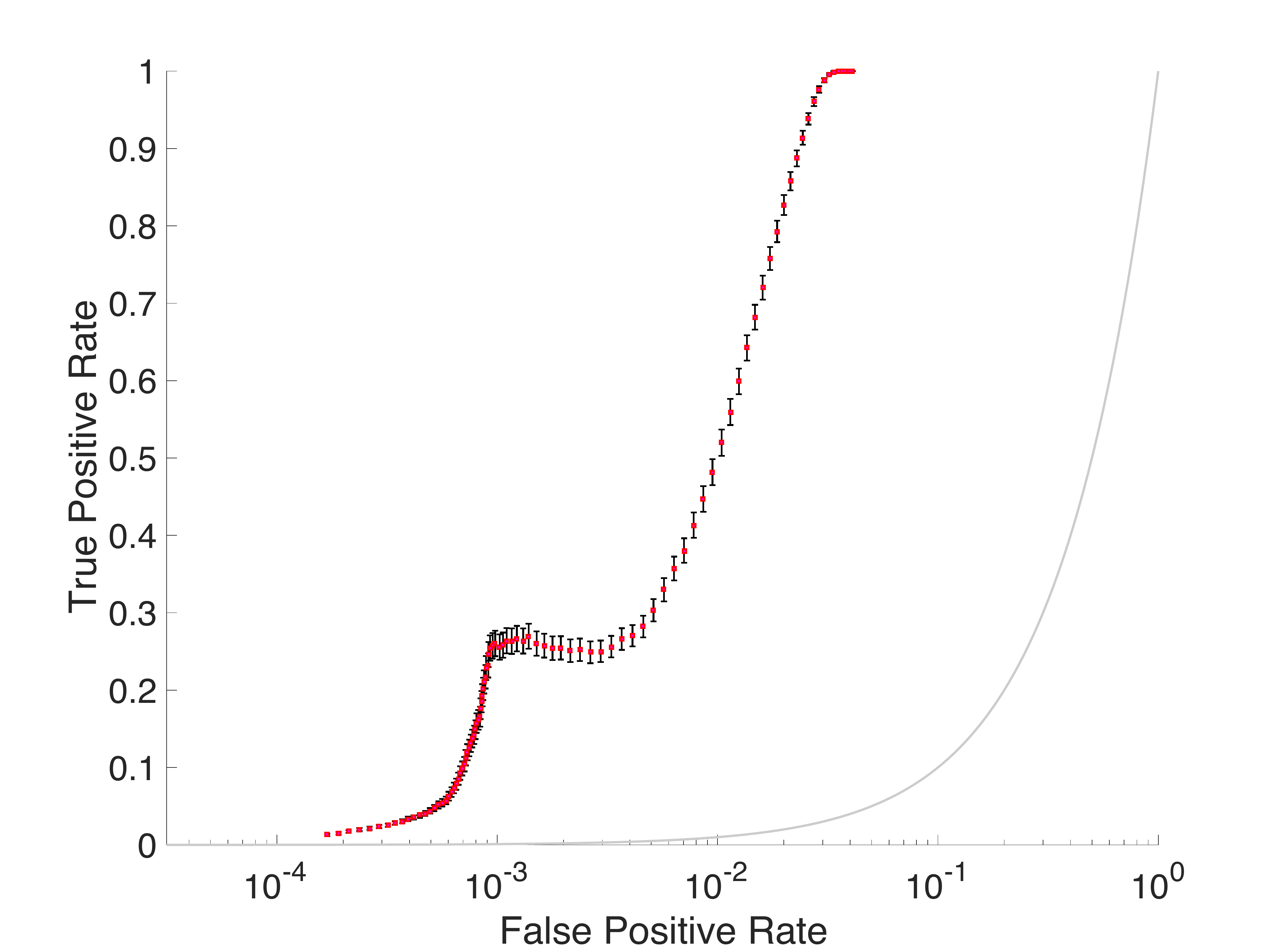}\\
        (b)
    \caption{ROC curves for the force-detection protocol based on computing derivatives of smoothed trajectory predictions, for two different impulse parameters: (a) $s=10$kHz, $w=0.015$ms; (b) $s=15$kHz, $w=0.02$ms. The gray (solid) curve shows the trivial ROC where the decision about whether each detected peak is a pulse or not is made randomly (with probability $1/2$). \label{fig:ROC}}
\end{figure}

As an alternative, we use an autocorrelation filter to further smooth the second order derivative signal in order to determine the time instant when an impulse-force arrives. We set up the following kernel function, \[\varphi(t) = \begin{cases}
1 & 0\textrm{ms} \le t < 0.03 \textrm{ms} \\
-1 & 0.03\textrm{ms}  \le t < 0.06 \textrm{ms} ,
\end{cases}\] and convolve this kernel with the $\dis \frac{\d^2\<x\>_S}{\d t^2}$ signal. The processed signal, $\dis \varphi(t)*\frac{\d^2\<x\>_S}{\d t^2}$, has very sharp peaks at discontinuities of $\dis \frac{\d^2\<x\>_S}{\d t^2}$, and suppressed noise. To set the detection threshold, we find the largest peak in the processed signal, $h$,  and set $\alpha h$ as the threshold for detecting the remaining peaks. 
In practice we find that $\alpha=0.5$ achieves a good balance between detection efficiency and specificity. In fact for pulses that are strong and wide (\eg $s>20$kHz and $w>0.03$ms) we find that the protocol has a true positive rate of almost one and a false positive rate of almost zero. In order to understand the effect of the choice of $\alpha$ for shorter, weaker pulses we plot receiver operating characteristic (ROC) curves for the protocol for two sample cases (that are representative of performance on short, weak pulses) in \cref{fig:ROC}. As can be seen from these ROC curves, the tradeoff between true positive rate and false positive rate is reasonable, even for short, weak pulses.

\end{document}